# MODELLING GESTURE BASED UBIQUITOUS APPLICATIONS

Kurien Zacharia[1], Eldo P. Elias[2], Surekha Mariam Varghese[3]

Department of Computer Science and Engineering, M. A. College of Engineering, Kothamangalam, Kerala, India
`kurienzach@gmail.com; eldope@gmail.com; surekha@mace.ac.in`

## ABSTRACT

*A cost effective, gesture based modelling technique called Virtual Interactive Prototyping (VIP) is described in this paper. Prototyping is implemented by projecting a virtual model of the equipment to be prototyped. Users can interact with the virtual model like the original working equipment. For capturing and tracking the user interactions with the model image and sound processing techniques are used. VIP is a flexible and interactive prototyping method that has much application in ubiquitous computing environments. Different commercial as well as socio-economic applications and extension to interactive advertising of VIP are also discussed.*

## KEYWORDS

*Rapid Prototyping, Virtual Reality, Ubiquitous Computing*

## 1. INTRODUCTION

Prototyping is the process of modelling design ideas prior to the implementation of the final product. This helps to collect variety opinions, and catch design flaws at an early stage. Virtual Interactive Prototyping (VIP) allows creating and testing prototypes easily by using computer simulated objects displayed on to an actual physical model using a projector [1]. Augmented reality concepts are used to help the users to display and work on a computer constructed model by projecting the interaction panel over a real physical model without actually assembling the components. User Interactions with the physical model is tracked through use of computer vision and sound techniques.

Even though the methods such as prototyping workbench allow creating hardware interface with relative ease, the cost factor makes them beyond the reach of many industries. VIP is a fast, flexible and interactive prototyping method and has much application in ubiquitous computing environments. In ubiquitous environments, modelling and evaluation of applications in an early stage is very important. Unavailability of details about user requirements and implementation makes modelling difficult. The degree to which a ubiquitous computing application can be useful depends on the ability of the designer to understand and anticipate the user needs. The acceptance level of any ubiquitous computing application depends actually on the effectiveness of the prototyping method. Simulation and modelling using immersive video techniques often require very high quality equipments which are very costly. VIP provides a cost effective prototyping technique for such applications. Also by effectively conveying the design to team members, VIP facilitate for iterative development and improvement

Virtual Interactive Prototyping is a virtual reality prototyping method based on a project Sixth Sense[2] and an existing prototyping method called Display Objects[3]. A physical model is first made from Cardboard or Styrofoam in the actual size and shape of the equipment to be prototyped. The components of the equipment (viz. screen, buttons etc.) are then projected from a projector connected to the computer into the physical model. A web camera tracks the user gestures as well as the physical model and reacts appropriately. A microphone is attached to the virtual prototype to detect sound produced during user taps. Computer Vision techniques and Sound processing techniques are used to detect user interactions so that the user can touch and interact with the prototype like actual working equipment.

## 2. RELATED WORK

Sixth Sense[2] is an innovative computing method that uses user gestures to perform day to day activities. Sixth Sense uses a camera, a mini projector, microphone, and color markers to track fingers. It projects information onto surfaces, walls, and physical objects around us, and lets us interact with the projected information through natural hand gestures, arm movements, or our interaction with the object itself. 'SixthSense' attempts to free information from its confines by seamlessly integrating it with reality, and thus making the entire world the user's computer. Sixth Sense actually changes the way we interact with computers and is used for a variety of interesting applications such as checking if a flight is delayed or not just by showing the ticket, automatically getting the reviews of books just by showing its Cover etc. Sixth sense uses the camera to locate the position of the color markers. Markers with distinct color are placed in each of the fingers. User finger movements are tracked by locating the position of the colour markers placed at the fingers.

Several studies have focused on the use of virtual interactive techniques in modelling and evaluation.. Avrahami, D. and Hudson used interactive techniques for product design [5]. Use of static virtual artefacts has been proposed for the analysis of team collaboration in virtual environments [6]. The display method uses projections to render a realistic model of the object according to the users viewing position. Also [7] (Maldovan, K., Messner, J.I. and Faddoul, M., (2006)) used a virtual mockup process for reviewing designs of physical structures such court rooms, restaurants etc. Pushpendra et. al. proposed an approach using immersive video with surround sound and a simulated infrastructure to create a realistic simulation of a ubiquitous environment in the software design and development office[8].

This paper proposes a system combining the technology used in the sixth sense system to implement the prototyping method as proposed in the display objects method for ubiquitous applications. To improve the effectiveness of the model, touch sensing with sound tracking of user actions is incorporated. A microphone is attached to the surface of the physical model, and the change in input level from the touch of the finger is detected. The result is a very low cost system with minimal setup or calibration time without compromising on the usability and performance of the system.

## 3. IMPLEMENTATION

The virtual modelling is implemented with the help of common image and sound processing techniques. For image processing, EmguCV, a C# wrapper for the OpenCV image processing library developed by Intel is used. The Open Source Computer Vision Library (OpenCV)[9] is a comprehensive computer vision library and machine learning written in C++ and C with additional Python and Java interfaces. It officially supports Linux, Mac OS, Windows, Android etc.

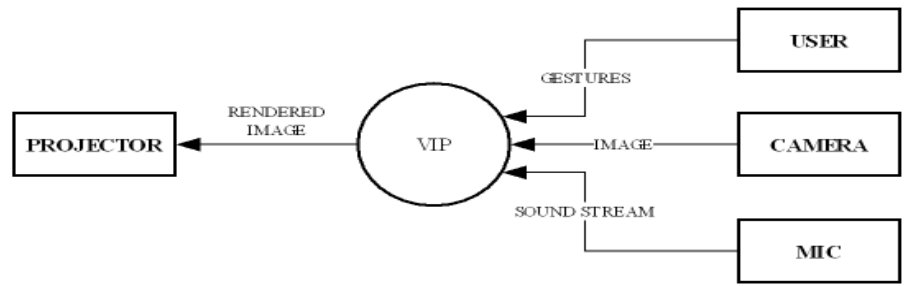

Figure 1. Level 0 Data Flow Diagram

The programming language used is C#.net as it facilitates fast development. For audio capturing and processing BASS audio processing library is used. All the core libraries that are used for implementing VIP are open source and platform independent. However VIP is implemented in Windows platform.

The Virtual Interactive Prototyping consists of the following main modules.

### 3.1. Marker Object tracking

This module deals with image processing and sound processing associated with finding the location of colour markers and also detecting the sound coming from the microphones to detect if the user has touched on the surface of the palette or display object.

#### 3.1.1. Segmentation

Segmentation is process of separating out a particular colour from an image[10][11]. To track the color marker we use a HSV color space based segmentation. The end result is a binary image with a particular color in white and all other colors in black.

**Algorithm for Segmentation**

```
Get Image From The Camera Capture
Get Segmentation Parameters From Parameter Settings(Hue_Min, Hue_Max, Value_Min,
Value_Max, Sat_Min, Sat_Max)
Generate New Blank Image With Size Img.Width, Img.Height
For Each Pixel In The Image [X,Y] Do
Get Color Data Of Pixel
If Color Data Lies B/W Segmentation Parameters
Set Pixel Of New Image [X,Y] = New Color(255)
Else  Set Pixel Of New Image [X,Y] = New Color(0)
```

### 3.1.2. Pointer Position Identification

Once the binary image from segmentation is obtained it is then fitted with a minimum area rectangle through the edges of the binary image. The centre point of this rectangle is then treated as the position of the finger or color marker.

### 3.1.3.  Click Detection and Generation

The sound stream from the microphone attached is captured continuously. A band pass filter is applied to the raw input stream to avoid unwanted noise. The level of resulting stream is then compared with a threshold value. If the level exceeds the threshold then it is treated as a click, a system call is made to create a mouse click at the current position of the color marker. The click detection is done using [user32.dll] library.

### 3.1.3 Display Object Tracking

Object tracking is similar to marker/finger tracking. It involves edge detection and click detection. Click Detection and Generation part is same for finger tracking and object tracking.

### 3.1. Display Object Tracking

Object tracking is similar to marker/finger tracking. It involves edge detection and click detection. Click Detection and Generation part is same for finger tracking and object tracking.

### 3.2.1. Edge Detection

In order to find the position of the display object, edges in the object are identified. Any edge detection methods such as Sobel operator, Prewitt operator, Robert operator, Laplace operator, LOG operator, Canny operator and so on can be used for this purpose.[12]. But gradient-based algorithms are very sensitive to noise. In a noisy environment, to distinguish valid image contents from visual artifacts introduced by noise an adaptive edge-detection algorithm is necessary. Hence Canny Edge detection algorithm, which works on gray scale images, is used for Edge detection. The algorithm is computationally more expensive compared to Sobel, Prewitt and Robert's operator. However, under almost all scenarios, the algorithm performs better than all the previously mentioned operators. In Canny's algorithm, if a pixel has a gradient larger than the upper threshold, then it is accepted as an edge pixel; if a pixel is below the lower threshold, it is rejected. If the pixel's gradient is between the thresholds, then it will be accepted only if it is connected to a pixel that is above the high threshold. The performance of the Canny algorithm depends heavily on the standard deviation for the Gaussian filter, and the threshold values, 'T1' and 'T2'. The size of the Gaussian filter is controlled by the standard deviation [13]. For noisy and larger images, more blurring is necessary. The user can tailor the algorithm by adjusting these parameters to adapt to different environments. Canny recommended a ratio of high:low threshold between 2:1 and 3:1. On trial error basis a threshold of 2.5:1 was selected in the implementation.

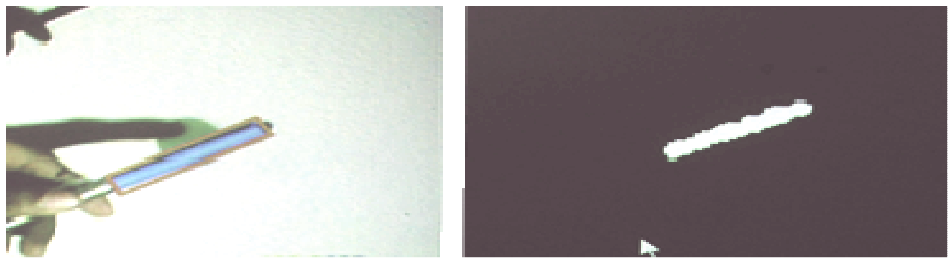

Figure 2. Color marker and Result of segmentation

### 3.2. Object Rendering

The rendering module involves placing the transformed composite image at the surface of the display panel projected on the position obtained from the Display Object Tracking module. Rendering involves Normalisation, Transformation and Display of the images/Objects. Image normalization is applied as a pre-processing step. The basic idea of image transformation is to convert the given image into a standard form which is independent of the distortions applied to the image. All the elements to be projected are first normalized into JPEG format. Image transforms are used to transform images into various shapes so that they perfectly fit inside the display object and move with them. The various transforms possible are scaling, rotation, distortion. A drawing method is used to create a composite image of the device from all the component images and then an affine transformation method is used to fit the image into the display object.

## 4. MODELLING THE APPLICATION

VIP can be used in any ubiquitous environment, where the user movements can be traced by placing color markers. Interactions involved in driving a car or movement of a customer in a super market etc. can be tracked by suitably placing the color markers in the body parts of the user. Combination of color markers and user gestures can be used to interact with the model. Each interaction triggers a predefined set of specified actions.

VIP is useful for demonstration and learning. The users of sophisticated equipments may be given demonstrations on how the equipment should be used and handled. A demo prototype was implemented for a mobile platform. The details of implementation are discussed in the remaining sections 5, 6 and 7. Section 8 and 9 describes the application of VIP in demonstration and interactive advertisements.

## 5. CONSTRUCTING THE PROTOTYPE

Construction of the prototype involves four steps. Creating a virtual palette on the monitor, drag and drop all the necessary components from a virtual palette into the panel on the virtual prototype using hand gestures, program the elements in the panel and interact with the constructed model. The components can be rearranged, moved and resized inside the Panel. After all the required components are set up, the user can then start the software simulation of the product. The user can use the model like the real equipment, in the way the programmer has programmed it. The user can have both the touch and feel of a real hardware prototype as well as all the interactions.

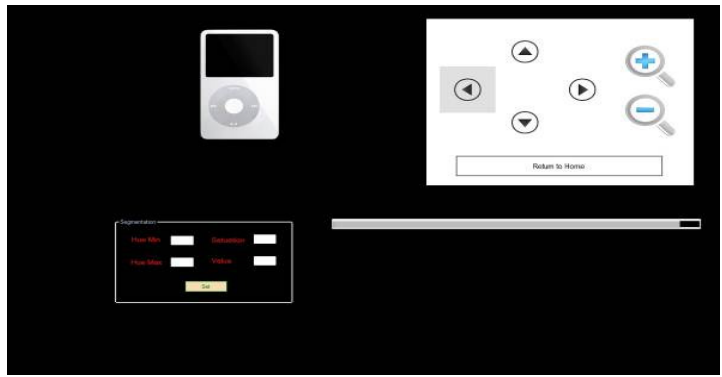

Figure 3. Developing the Virtual Model

## 6. COMPONENT PROGRAMMING

Two types of interactions are possible with the virtual model. During edit mode components can be included, repositioned, resized and programmed. Programming involves specifying input and output elements associated with each of the panel elements. After placing the components/ panel elements on the model, a script generates a Quartz file for the panel. The programming can be performed in three ways: on a computer, on the Palette, or on the physical model itself. Edit mode can be initiated with the corresponding button click or gesture. Connections can be established between input elements, components and output elements by dragging appropriate connectors between each pair using a Quartz Composer on the system. For example, a screen element, as shown in Figure 3 can be scripted to associate with the contents of a movie file connected to one

The International Journal of Multimedia & Its Applications (IJMA) Vol.3, No.4, November 2011

of its inlets. Simple interactive behaviors, such as starting, stopping or scrolling through the movie file can be linked with the screen element. More complicated behaviors are made possible by dragging Quartz Composer subpatches from the palette onto the physical model, and connecting inlets and outlets. These subpatches are programmed on the computer, rather than on the model. Because subpatches are a regular part of Quartz Composer, they may contain Objective-C code to further extend behaviors ranging from video feeds, to screen captures, Flash content, live webpages, etc.

## 7. INTERACTION TECHNIQUES

The workbench incorporates both one-handed and two handed techniques so that the user can interact with the model with one hand while browsing interaction elements with the other. The non-dominant hand is typically used to hold the physical model, while the dominant hand picks up interaction elements from the Palette. The Palette can be held or placed on a surface if hands are occupied with a physical model. Interaction elements are then pasted and placed onto the physical model using simple tapping and dragging techniques. To allow recognition of finger action by the system, fingers need to be augmented with a Vicon marker. Typically, only one or two fingers need to be enhanced in this way. Gestures allow fine-grained actions on the physical model and Palette.

### 7.1. Scanning

The user can simply move the finger over the surface of a physical model or Palette. This does not result in an action until the surface is touched.

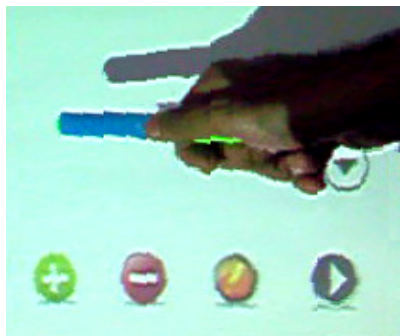

Figure 4. Use of color marker for interactions

### 7.2. Selecting

Users can select elements from the Palette by pointing the element in the palette and tapping once.

### 7.3. Placing

Users can place the selected elements from the Palette on the physical model.. Selection of the element makes the element to be copied to a virtual clipboard at the tip of the finger.  A subsequent tap on the physical model places the element on the particular position of the physical model where the tap was performed.

### 7.4. Dragging

Users can move elements on the surface of the physical model by touching them while in edit mode, and then dragging their finger on the surface of the physical model.

### 7.5. Locking

Element placed on the panel should be fixed to a particular position by locking it. After finalizing the position, the users can lock interaction elements on the panel by tapping the "Lock" button on the Palette.

### 7.6. Clicking

Simulation of a button click or element action in the panel involves two steps. First the element placed on the panel should be fixed to a particular position by locking it and then the action associated with the particular element can be played. Once the element is locked, any tap on the element is interpreted as a click on the element that executes its associated functionality. This provides for functional interaction with the elements on the surface of the panel, allowing users, for example, to press buttons that trigger some visual action or play some audio files.

### 7.7. Resizing

Resizing of Interaction elements are performed with the help of Pinch and Release gestures. The Pinch gesture is initiated by bringing a thumb and index finger into contact with the extremities of a target element, and then drawing the thumb and index finger together. The Release gesture is performed by reversing this action.

### 7.8. Wiping

Users can move through layers of content on the surface of the panel by performing a wipe gesture, from the bottom left corner of the physical model upward. When editing objects, this is used to reveal the Quartz Composer patch that underlies the functionality of the panel. Layers can be closed by an inverse wipe.

## 8. DEMONSTRATION PROTOTYPING

VIP was used to simulate the working of an ATM machine. An augmented display is arranged for naïve users of ATM to help and direct them. The virtual prototype constructed is presented in figure 5. Many users will find it difficult to use new ATM machines with additional functionalities. Due to security protection, consequent erroneous/wrong/incorrect usage will lead to temporary blocking of the account. During such situations, instead of blocking the account, the users can be provided with a virtual model. This virtual model will help the users to learn the interactions with the ATM.

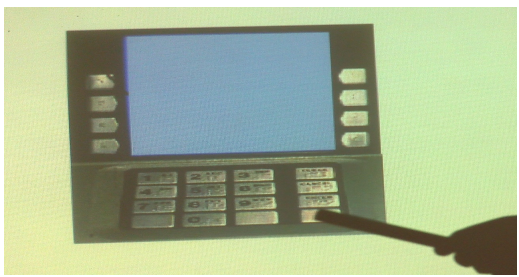

Figure 5. Interactions with Virtual ATM model

## 9. FUTURE WORK

Much of the previous research on context aware computing has sought to develop systems that could detect and interpret contextual characteristics of a user environment. But very less efforts were taken on improving the usability of these systems. Interactive advertising is a context aware application that traces user movements in front of the system and displays advertisements and warnings according to their gestures and movements. It is a highly effective branding and marketing tool that helps in capturing the attention of target audiences and attract passerby's for improving advertising effectiveness. It is an efficient means of communication with consumers for promotion of products, brands and services. VIP can be used for interactive advertising by placing the camera near a large display in a public place. The advertisements can be displayed only if there are movements in front of the display. Other than peak times, automatic off feature saves much power. Crowd identification and identification of kids in the group can initiate group specific advertisements. Alice et. al used Principal Component analysis of face image for gender determination[14]. Gender specific addressing can be incorporated by proper analysis of the head and facial parts of the user using similar methods.

Now a days the main medium of interactive advertisement is the Internet. But when it is used for study of the methods, the aspects figured out will be partial, since a large group of target audience is not considered in Internet. VIP overcomes this drawback by incorporating general public into the study.

The various aspects of Internet advertising like banner ads, sponsorship, hyperlinks are controlled by the advertiser. These aspects include structural elements, such as ad types, formats and features. Consumers have less control over the structure of the interactive ads. Interactive advertising using VIP also suffers from the same problems but since VIP is used in a public place, it can consider the local taste of consumers. Hence it can be termed as more user oriented.

## 10. CONCLUSION

Revising and modelling of applications and equipments is an urgent need of Today's industry. Many of the existing methods are slumpy and often require a near full-scale deployment. In this paper, a rapid prototyping workbench called VIP which is helpful for designing new devices in such a context is discussed. A successful example elaborating on the effective use of the workbench in prototyping a mobile phone and ATM machine is demonstrated. Main advantage of the method is the ability for real time modifications of the prototype, which is not possible on a real prototype. Since all the libraries used for implementation are open source, the workbench can be made available as Open Source.

Ubiquitous applications need natural interfaces that facilitate smooth interactions between humans and the system. The applications should be designed to be context-aware so that they adapt their behavior according to the information sensed from the environment. Complexity and diversity of interactions and need for context awareness complicates the design of the ubiquitous applications. There are numerous applications in which full picture is not available when the devices are designed. It is important to explore how the valued interactions can be improved, extended and augmented rather than entirely supplanted. VIP is a simple, cost effective method to model ubiquitous applications where visual interactions are applicable.

Another important property is the reusability of components and interface. Also the extension of VIP to interactive advertising is a promising area.

## Authors

Kurien Zacharia completed his B-Tech in Computer Science and Engineering from M.A. College of Engineering, Kothamangalam in 2011. He has presented many papers in national level competitions. His areas of research are Data Structures, In-memory d atabases, Computer Graphics and Virtual Reality

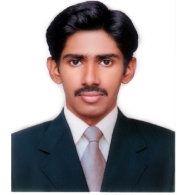

Eldo P Elias completed his BE in Computer Science and Engineering from Sri Ramakrishna Engineering College, Coimbatore in 2003. He has more th an 7 years of teaching experience at M.A College of Engineering, Kothamangalam. His areas of research are Networking, Operating Systems and Computer Security.

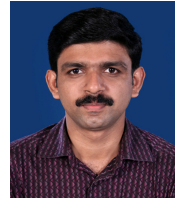

Surekha Mariam Varghese is currently heading the Department of Computer Science and Engineering, M.A. College of Engineering, Kothamangalam, Kerala, India. She received her B-Tech Degree in Computer Science and Engineering in 1990 from College of Engineering, Trivandrum affiliated to Kerala University and M-Tech in Computer and Information Sciences from Cochin University of Science and Technology, Kochi in 1996. She obtained Ph.D in Computer Security from Cochin U niversity of Science and Technology, Kochi in 2009. She has around 20 years of teaching and research experience in various institutions in India. Her research interests include Network Security, Database Management, Data Structures and Algorithms, Operating Systems and Distributed Computing. She has published 6 papers in international journals and international conference proceedings. She has served as reviewer, committee member and session chair for many international conferences and journals.

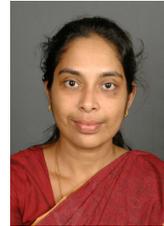